\documentclass[aps,prl,twocolumn,superscriptaddress,showpacs,amsmath,amssymb]{revtex4}

\usepackage{amsmath}
\usepackage{graphicx}% Include figure files
\usepackage{dcolumn}% Align table columns on decimal point
\usepackage{bm}% bold math
\usepackage{xspace}
\usepackage{multirow}
\usepackage{booktabs}
\usepackage{longtable}
\usepackage{color}
\newcommand{\phid}{\ensuremath{\phi}\xspace}
\newcommand{\phidpi}{\ensuremath{\phid \pi^+}\xspace}
\newcommand{\phidpim}{\ensuremath{\phid \pi^-}\xspace}

\newcommand{\KS}{\ensuremath{\bar{K}^{*0}}\xspace}

\newcommand{\KSd}{\ensuremath{\bar{K}^{*0}}\xspace}
\newcommand{\KSdK}{\ensuremath{\KSd K^+}\xspace}
\newcommand{\KSdKm}{\ensuremath{K^{*0} K^-}\xspace}

\newcommand{\Dminus}{\ensuremath{D^-}\xspace}
\newcommand{\Dplus}{\ensuremath{D^+}\xspace}
\newcommand{\DSplus}{\ensuremath{D^{*+}}\xspace}
\newcommand{\DSminus}{\ensuremath{D^{*-}}\xspace}

\newcommand{\Kpipi}{\ensuremath{K^+ \pi^- \pi^-}\xspace}
\newcommand{\DKpipi}{\ensuremath{\Dplus \rightarrow K^- \pi^+ \pi^+}\xspace}

\newcommand{\DSDsgammapi}{\ensuremath{\DSplus \rightarrow \Dplus \gamma/\pi^{0}}\xspace}

\newcommand{\Ds}{\ensuremath{D_s^\pm}\xspace}
\newcommand{\Dsmp}{\ensuremath{D_s^{\mp}}\xspace}
\newcommand{\Dsplus}{\ensuremath{D_s^+}\xspace}
\newcommand{\Dsminus}{\ensuremath{D_s^-}\xspace}
\newcommand{\DsSplus}{\ensuremath{D_s^{*+}}\xspace}
\newcommand{\DsSminus}{\ensuremath{D_s^{*-}}\xspace}
\newcommand{\DsSpm}{\ensuremath{D_s^{*\pm}}\xspace}

\newcommand{\DsplusInc}{\ensuremath{D_s^{(*)+}}\xspace}
\newcommand{\DsminusInc}{\ensuremath{D_s^{(*)-}}\xspace}
\newcommand{\DsKKpi}{\ensuremath{\Dsplus \rightarrow K^+ K^- \pi^+}\xspace}
\newcommand{\DsSDsgamma}{\ensuremath{\DsSplus \rightarrow \Dsplus \gamma}\xspace}
\newcommand{\DsSDspi}{\ensuremath{\DsSplus \rightarrow \Dsplus \pi^0}\xspace}

\newcommand{\Bzero}{\ensuremath{B^0}\xspace}
\newcommand{\BDDs}{\ensuremath{\Bzero \rightarrow \Dsplus\Dminus}\xspace}

\newcommand{\BDphidpi}{\ensuremath{\Bzero \rightarrow \Dsplus(\rightarrow \phidpi)\Dminus}\xspace}

\newcommand{\Bs}{\ensuremath{B_s^0}\xspace}
\newcommand{\Bsbar}{\ensuremath{\bar{B}_s^0}\xspace}
\newcommand{\BsL}{\ensuremath{B_{sL}^0}\xspace}
\newcommand{\BsH}{\ensuremath{B_{sH}^0}\xspace}
\newcommand{\BsDsDs}{\ensuremath{\Bs \rightarrow \Dsplus\Dsminus}\xspace}
\newcommand{\BsDsSDs}{\ensuremath{\Bs \rightarrow \DsSpm\Dsmp}\xspace}
\newcommand{\BsDsSDsS}{\ensuremath{\Bs \rightarrow \DsSplus\DsSminus}\xspace}
\newcommand{\BsDsDsInc}{\ensuremath{\Bs \rightarrow \DsplusInc\DsminusInc}\xspace}

\newcommand{\DGs}{\ensuremath{\Delta\Gamma_s}\xspace}

\newcommand{\Br}[1]{\ensuremath{\mathcal{B}(#1)}\xspace}
\newcommand{\fDsDs}{\ensuremath{f_{D_s D_s}}\xspace}
\newcommand{\fDsSDs}{\ensuremath{f_{D_s^* D_s}}\xspace}
\newcommand{\fDsSDsS}{\ensuremath{f_{D_s^* D_s^*}}\xspace}
\newcommand{\fDsDsInc}{\ensuremath{f_{D_s^{(*)} D_s^{(*)}}}\xspace}

\newcommand{\fsfd}{\ensuremath{\frac{f_s}{f_d}}\xspace}
\newcommand{\fL}{\ensuremath{f_{L}}\xspace}
\newcommand{\fcpodd}{\ensuremath{f_{CP-}}\xspace}

\begin{document}
%\linenumbers

% remove the space for publication
%\vspace*{1.5cm}

\title{Measurement of \BsDsDsInc Branching Ratios} %\\ \vspace*{2.0cm}}

\affiliation{Institute of Physics, Academia Sinica, Taipei, Taiwan 11529, Republic of China}
\affiliation{Argonne National Laboratory, Argonne, Illinois 60439, USA}
\affiliation{University of Athens, 157 71 Athens, Greece}
\affiliation{Institut de Fisica d'Altes Energies, ICREA, Universitat Autonoma de Barcelona, E-08193, Bellaterra (Barcelona), Spain}
\affiliation{Baylor University, Waco, Texas 76798, USA}
\affiliation{Istituto Nazionale di Fisica Nucleare Bologna, $^{ee}$University of Bologna, I-40127 Bologna, Italy}
\affiliation{University of California, Davis, Davis, California 95616, USA}
\affiliation{University of California, Los Angeles, Los Angeles, California 90024, USA}
\affiliation{Instituto de Fisica de Cantabria, CSIC-University of Cantabria, 39005 Santander, Spain}
\affiliation{Carnegie Mellon University, Pittsburgh, Pennsylvania 15213, USA}
\affiliation{Enrico Fermi Institute, University of Chicago, Chicago, Illinois 60637, USA}
\affiliation{Comenius University, 842 48 Bratislava, Slovakia; Institute of Experimental Physics, 040 01 Kosice, Slovakia}
\affiliation{Joint Institute for Nuclear Research, RU-141980 Dubna, Russia}
\affiliation{Duke University, Durham, North Carolina 27708, USA}
\affiliation{Fermi National Accelerator Laboratory, Batavia, Illinois 60510, USA}
\affiliation{University of Florida, Gainesville, Florida 32611, USA}
\affiliation{Laboratori Nazionali di Frascati, Istituto Nazionale di Fisica Nucleare, I-00044 Frascati, Italy}
\affiliation{University of Geneva, CH-1211 Geneva 4, Switzerland}
\affiliation{Glasgow University, Glasgow G12 8QQ, United Kingdom}
\affiliation{Harvard University, Cambridge, Massachusetts 02138, USA}
\affiliation{Division of High Energy Physics, Department of Physics, University of Helsinki and Helsinki Institute of Physics, FIN-00014, Helsinki, Finland}
\affiliation{University of Illinois, Urbana, Illinois 61801, USA}
\affiliation{The Johns Hopkins University, Baltimore, Maryland 21218, USA}
\affiliation{Institut f\"{u}r Experimentelle Kernphysik, Karlsruhe Institute of Technology, D-76131 Karlsruhe, Germany}
\affiliation{Center for High Energy Physics: Kyungpook National University, Daegu 702-701, Korea; Seoul National University, Seoul 151-742, Korea; Sungkyunkwan University, Suwon 440-746, Korea; Korea Institute of Science and Technology Information, Daejeon 305-806, Korea; Chonnam National University, Gwangju 500-757, Korea; Chonbuk National University, Jeonju 561-756, Korea}
\affiliation{Ernest Orlando Lawrence Berkeley National Laboratory, Berkeley, California 94720, USA}
\affiliation{University of Liverpool, Liverpool L69 7ZE, United Kingdom}
\affiliation{University College London, London WC1E 6BT, United Kingdom}
\affiliation{Centro de Investigaciones Energeticas Medioambientales y Tecnologicas, E-28040 Madrid, Spain}
\affiliation{Massachusetts Institute of Technology, Cambridge, Massachusetts 02139, USA}
\affiliation{Institute of Particle Physics: McGill University, Montr\'{e}al, Qu\'{e}bec, Canada H3A~2T8; Simon Fraser University, Burnaby, British Columbia, Canada V5A~1S6; University of Toronto, Toronto, Ontario, Canada M5S~1A7; and TRIUMF, Vancouver, British Columbia, Canada V6T~2A3}
\affiliation{University of Michigan, Ann Arbor, Michigan 48109, USA}
\affiliation{Michigan State University, East Lansing, Michigan 48824, USA}
\affiliation{Institution for Theoretical and Experimental Physics, ITEP, Moscow 117259, Russia}
\affiliation{University of New Mexico, Albuquerque, New Mexico 87131, USA}
\affiliation{The Ohio State University, Columbus, Ohio 43210, USA}
\affiliation{Okayama University, Okayama 700-8530, Japan}
\affiliation{Osaka City University, Osaka 588, Japan}
\affiliation{University of Oxford, Oxford OX1 3RH, United Kingdom}
\affiliation{Istituto Nazionale di Fisica Nucleare, Sezione di Padova-Trento, $^{ff}$University of Padova, I-35131 Padova, Italy}
\affiliation{University of Pennsylvania, Philadelphia, Pennsylvania 19104, USA}
\affiliation{Istituto Nazionale di Fisica Nucleare Pisa, $^{gg}$University of Pisa, $^{hh}$University of Siena and $^{ii}$Scuola Normale Superiore, I-56127 Pisa, Italy}
\affiliation{University of Pittsburgh, Pittsburgh, Pennsylvania 15260, USA}
\affiliation{Purdue University, West Lafayette, Indiana 47907, USA}
\affiliation{University of Rochester, Rochester, New York 14627, USA}
\affiliation{The Rockefeller University, New York, New York 10065, USA}
\affiliation{Istituto Nazionale di Fisica Nucleare, Sezione di Roma 1, $^{jj}$Sapienza Universit\`{a} di Roma, I-00185 Roma, Italy}
\affiliation{Rutgers University, Piscataway, New Jersey 08855, USA}
\affiliation{Texas A\&M University, College Station, Texas 77843, USA}
\affiliation{Istituto Nazionale di Fisica Nucleare Trieste/Udine, I-34100 Trieste, $^{kk}$University of Udine, I-33100 Udine, Italy}
\affiliation{University of Tsukuba, Tsukuba, Ibaraki 305, Japan}
\affiliation{Tufts University, Medford, Massachusetts 02155, USA}
\affiliation{University of Virginia, Charlottesville, Virginia 22906, USA}
\affiliation{Waseda University, Tokyo 169, Japan}
\affiliation{Wayne State University, Detroit, Michigan 48201, USA}
\affiliation{University of Wisconsin, Madison, Wisconsin 53706, USA}
\affiliation{Yale University, New Haven, Connecticut 06520, USA}

\author{T.~Aaltonen}
\affiliation{Division of High Energy Physics, Department of Physics, University of Helsinki and Helsinki Institute of Physics, FIN-00014, Helsinki, Finland}
\author{B.~\'{A}lvarez~Gonz\'{a}lez$^z$}
\affiliation{Instituto de Fisica de Cantabria, CSIC-University of Cantabria, 39005 Santander, Spain}
\author{S.~Amerio}
\affiliation{Istituto Nazionale di Fisica Nucleare, Sezione di Padova-Trento, $^{ff}$University of Padova, I-35131 Padova, Italy}
\author{D.~Amidei}
\affiliation{University of Michigan, Ann Arbor, Michigan 48109, USA}
\author{A.~Anastassov$^x$}
\affiliation{Fermi National Accelerator Laboratory, Batavia, Illinois 60510, USA}
\author{A.~Annovi}
\affiliation{Laboratori Nazionali di Frascati, Istituto Nazionale di Fisica Nucleare, I-00044 Frascati, Italy}
\author{J.~Antos}
\affiliation{Comenius University, 842 48 Bratislava, Slovakia; Institute of Experimental Physics, 040 01 Kosice, Slovakia}
\author{G.~Apollinari}
\affiliation{Fermi National Accelerator Laboratory, Batavia, Illinois 60510, USA}
\author{J.A.~Appel}
\affiliation{Fermi National Accelerator Laboratory, Batavia, Illinois 60510, USA}
\author{T.~Arisawa}
\affiliation{Waseda University, Tokyo 169, Japan}
% ...

\collaboration{CDF Collaboration\footnote{With visitors from
$^a$Istituto Nazionale di Fisica Nucleare, Sezione di Cagliari, 09042 Monserrato (Cagliari), Italy,
$^b$University of CA Irvine, Irvine, CA 92697, USA,
$^c$University of CA Santa Barbara, Santa Barbara, CA 93106, USA,
$^d$University of CA Santa Cruz, Santa Cruz, CA 95064, USA,
$^e$Institute of Physics, Academy of Sciences of the Czech Republic, Czech Republic,
$^f$CERN, CH-1211 Geneva, Switzerland,
$^g$Cornell University, Ithaca, NY 14853, USA,
$^h$University of Cyprus, Nicosia CY-1678, Cyprus,
$^i$Office of Science, U.S. Department of Energy, Washington, DC 20585, USA,
$^j$University College Dublin, Dublin 4, Ireland,
$^k$ETH, 8092 Zurich, Switzerland,
$^l$University of Fukui, Fukui City, Fukui Prefecture, Japan 910-0017,
$^m$Universidad Iberoamericana, Mexico D.F., Mexico,
$^n$University of Iowa, Iowa City, IA 52242, USA,
$^o$Kinki University, Higashi-Osaka City, Japan 577-8502,
$^p$Kansas State University, Manhattan, KS 66506, USA,
$^q$Ewha Womans University, Seoul, 120-750, Korea,
$^r$University of Manchester, Manchester M13 9PL, United Kingdom,
$^s$Queen Mary, University of London, London, E1 4NS, United Kingdom,
$^t$University of Melbourne, Victoria 3010, Australia,
$^u$Muons, Inc., Batavia, IL 60510, USA,
$^v$Nagasaki Institute of Applied Science, Nagasaki, Japan,
$^w$National Research Nuclear University, Moscow, Russia,
$^x$Northwestern University, Evanston, IL 60208, USA,
$^y$University of Notre Dame, Notre Dame, IN 46556, USA,
$^z$Universidad de Oviedo, E-33007 Oviedo, Spain,
$^{aa}$CNRS-IN2P3, Paris, F-75205 France,
$^{bb}$Texas Tech University, Lubbock, TX 79609, USA,
$^{cc}$Universidad Tecnica Federico Santa Maria, 110v Valparaiso, Chile,
$^{dd}$Yarmouk University, Irbid 211-63, Jordan,
$^{mm}$University of Warwick, Coventry CV4 7AL, United Kingdom,
}}
\noaffiliation

% \author{The CDF Collaboration}
%\affiliation{URL http://www-cdf.fnal.gov}

\date{\today}

\begin{abstract}
The decays \BsDsDsInc are reconstructed in a data sample corresponding to an integrated 
luminosity of 6.8~fb$^{-1}$ collected by the CDF II detector at the Tevatron $p\bar{p}$ collider.
All decay modes are observed with a significance of more than 10$\,\sigma$,
and we measure the \Bs production rate times 
\BsDsDsInc branching ratios relative to the normalization mode \BDDs to be 
$0.183 \pm 0.021 \pm 0.017$ for \BsDsDs,
$0.424 \pm 0.046 \pm 0.035$ for \BsDsSDs,
$0.654 \pm 0.072 \pm 0.065$ for \BsDsSDsS, and
$1.261 \pm 0.095 \pm 0.112$ for the inclusive decay \BsDsDsInc,
where the uncertainties are statistical and systematic.
These results are the most precise single measurements to date and provide important constraints 
for indirect searches for non-standard model physics in \Bs mixing.
\end{abstract}

% activate the following line for publication
% 13.25.Hw Decays of bottom mesons
% 14.40.Nd Bottom Mesons (|B|>0)
% 12.15.Ff Quark and lepton masses and mixing

\maketitle

% introduction
A \Bs meson can oscillate into its antiparticle via second order weak interaction transitions,
which make its time evolution sensitive to contributions from new physics processes.
Such contributions are not well constrained yet and might be responsible for the
deviation from the standard model reported in Ref.~\cite{Abazov:2011yk}.
The \Bs eigenstates with defined mass and lifetime, \BsL and \BsH, are linear combinations
of the \Bs and \Bsbar states and, in the standard model, correspond in good approximation
to the even and odd \textit{CP} eigenstates, respectively.
In the absence of substantial \textit{CP} violation, a sizable decay width difference between the light and heavy mass eigenstates,
$\DGs = \Gamma_{sL} - \Gamma_{sH}$, arises from the fact that decays to final states of 
definite \textit{CP} are only accessible by one of the mass eigenstates.
The dominant contribution to \DGs is believed to come from the \BsDsDsInc decays~\cite{ChargeConjugate},
which are predominantly \textit{CP}-even and saturate \DGs under certain theoretical 
assumptions~\cite{Aleksan:1993qp,Shifman:1987rj}, resulting in the relation
\begin{equation}
2\Br{\BsDsDsInc} \approx \frac{\DGs}{\Gamma_s + \DGs/2},
\label{eqn:DGs}
\end{equation}
where $\Gamma_s = (\Gamma_{sL} + \Gamma_{sH}) / 2$ \cite{Dunietz:2000cr}.
However, three-body modes may provide a significant contribution to \DGs~\cite{Chua:2011er}.

A finite value of \DGs improves the experimental sensitivity to \textit{CP} violation
because it allows one to distinguish the two mass eigenstates via their decay time distribution.
Furthermore, the \BsDsDsInc decays could be used in future to measure directly 
the lifetime of the \textit{CP}-even eigenstate, which would complement
the \textit{CP}-odd eigenstate lifetime measurement in $\Bs \rightarrow J/\psi f_0(980)$ decays~\cite{Aaltonen:2011nk}
and provide additional information in the search for new physics contributions to
\textit{CP} violation in the \Bs system.

% status
The \BsDsDsInc decay modes have been previously studied by 
the ALEPH, CDF, D0, and Belle collaborations~\cite{Barate:2000kd, Abulencia:2007zz, Abazov:2008ig, Esen:2010jq}.
The current world average branching ratios~\cite{Nakamura:2010zzi}, 
which do not yet include the latest preliminary Belle results~\cite{arXiv:1110.2099}, are
$\Br{\BsDsDs} = (1.04^{+0.29}_{-0.26})$\,\%, $\Br{\BsDsSDs} = (2.8 \pm 1.0)$\,\%,
$\Br{\BsDsSDsS} = (3.1 \pm 1.4)$\,\%, and $\Br{\BsDsDsInc} = (4.5 \pm 1.4)$\,\%.

% overview
In a data sample corresponding to an integrated luminosity of 6.8~fb$^{-1}$ recorded by the CDF~II
detector at the Tevatron $p\bar{p}$ collider we reconstruct \BsDsDsInc decays with \DsKKpi.
For the first time in this channel, the acceptance is calculated using a \Ds Dalitz model
instead of a simple two-body decay model.
The photon and the neutral pion from the \DsSDsgamma and \DsSDspi decays are not
reconstructed because of their low detection efficiency.
In a simultaneous fit to the reconstructed $B^0_{(s)}$ meson invariant mass spectra we measure
the \Bs production rate times \BsDsDsInc branching ratios relative to the normalization 
mode \BDDs
\begin{equation}
f_X = \fsfd \frac{\Br{\Bs \rightarrow X}}{\Br{\BDDs}},
\label{eqn:fX} 
\end{equation}
for $X = \Dsplus\Dsminus$, $\DsSpm\Dsmp$, $\DsSplus\DsSminus$, and the inclusive $\DsplusInc\DsminusInc$
where $f_s/f_d$ is the relative rate of produced \Bs to \Bzero mesons.

% detector
The components of the CDF II detector~\cite{Acosta:2004yw} most relevant for this analysis
are the tracking systems located inside a solenoid that provides a 1.4~T magnetic field.
Charged particles' trajectories (tracks) are reconstructed in layers of silicon-strip sensors 
located between radii of 1.5~cm and 28~cm from the beam line and an open-cell drift chamber (COT) with a radial
extension from 40 to 137~cm.
Tracks with a pseudorapidity $|\eta|\le 1.0$ pass the full radial extent of the COT.
Kaons and pions are statistically identified by measurements of the ionization energy loss in the COT and
information from the time-of-flight system located between the COT and the solenoid.
The events for this analysis are selected online by identifying pairs of tracks detected
in the COT and the silicon detector~\cite{Thomson:2002xp}.
Minimal requirements on the momenta and the displacement of the tracks and the reconstructed decay vertex 
from the primary vertex are imposed.

% reconstruction and selection
We reconstruct \DsKKpi and \DKpipi decays from combinations of three tracks with appropriate
charge and mass hypothesis assignments, fitted to a common vertex.
Because the \DsKKpi decay proceeds mainly via $\phi \pi^+$ and $\KS K^+$, we select
candidates with $1.005 < m(K^+ K^-) < 1.035$ GeV$/c^2$ 
and $0.837 < m(K^- \pi^+) < 0.947$ GeV$/c^2$, centered on the known $\phi$ and $K^{*0}$ masses, respectively.
According to the \DsKKpi Dalitz structure~\cite{:2009tr} this requirement has a signal acceptance 
of about 75\,\% while covering only 14\,\%
of the phase space and thus increasing the signal-to-background ratio.
In the following we will denote the selected $K^+ K^-$ and $K^- \pi^+$ combinations
as \phid and \KSd, respectively, since the dominant contributions come from these resonances.
However, we implicitly include contributions from other resonances and interference
effects when using these terms.

Pairs of $\Dsplus \rightarrow \phidpi$ or $\Dsplus \rightarrow \KSdK$ candidates
and $\Dsminus \rightarrow \phidpim$ candidates
are combined to form \Bs candidates and fitted to a common vertex.
Combinations where both charm mesons decay into a \KSd mode
are not considered because of the low signal-to-background ratio.
Candidate \Bzero mesons are reconstructed from $\Dsplus\Dminus$ combinations
where both \Dsplus decay modes are used.

To reject background-like events, requirements are placed on track quality variables, $B$ meson momentum, 
reconstructed $D$ meson masses, vertex fit qualities, and vertex displacement significances.
To further increase the signal purity, two artificial neural networks are used, one for candidates with a \KSd 
and one for candidates without.
To minimize the systematic uncertainty of the relative selection efficiency,
the same networks are applied to \Bs and \Bzero candidates, and only information
from the \Ds that is common to both $B$ meson decays is used.
The networks are trained on simulated signal events, described below, and on background 
events from the 5.45 to 6.5 GeV$/c^2$ $B$ mass sideband.
The input variables contain kinematic, lifetime, fit quality, and particle identification 
information. 
The $B$ vertex displacement significance in the transverse plane gives the largest
contribution to the discrimination power of both networks.
The selection criteria on the network outputs are chosen such that they maximize the significance
$\epsilon_\mathrm{MC}/\sqrt{N_\mathrm{data}}$, where $\epsilon_\mathrm{MC}$ is the \Bs selection efficiency
determined from simulation and $N_\mathrm{data}$ is the number of data events in the \Bs signal
window from 5.343 to 5.397 GeV$/c^2$.

About 6\,\% of the selected \BDphidpi candidates also fulfill the \Bs selection requirements,
where the assignment of a \Dminus daughter track is swapped from pion to kaon.
To avoid having the same event entering the fit multiple times, we reject each event that is reconstructed as \Bs candidates
from the \Bzero sample.
The cross-populations between the two \Bs modes and between the two \Bzero modes, respectively, are negligible.
The selected sample contains about 750 \Bs signal events.

% MC, acceptance, efficiency
Simulated events are used to determine the reconstruction and selection efficiency.
The $B^0_{(s)}$ mesons are generated according to the momentum spectrum measured in exclusive $B$ decays
and decayed to the considered final states with the \textsc{evtgen} package~\cite{Lange:2001uf}. 
For the $\Bs$ meson we assign the lifetime of the $\BsL$ eigenstate~\cite{Nakamura:2010zzi}
that coincides with the \textit{CP}-even eigenstate in the standard model.
For all the other long-lived charm and bottom mesons, the world 
average mean lifetimes~\cite{Nakamura:2010zzi} are used.
The \BsDsSDsS decay is a transition of a pseudoscalar to two vector mesons and its angular
distribution is described by three polarization amplitudes.
Since these amplitudes are unknown, we take the same longitudinal polarization 
as measured in $\Bzero \rightarrow \DSplus \DSminus$ decays~\cite{Aubert:2003jj} and a vanishing
\textit{CP}-odd component as default values.
The world average value~\cite{Nakamura:2010zzi} is used for the ratio of \DsSDsgamma to \DsSDspi decays.
The dynamics of the decay \DsKKpi is simulated according to the Dalitz structure measured by CLEO~\cite{:2009tr}.
The generated events are processed by a \textsc{geant3} based detector simulation~\cite{Brun:1978fy}
and the same reconstruction program as applied to real data events.

% fit
The relative branching ratios times production rate are determined in a simultaneous
extended unbinned maximum-likelihood fit to the $(\phidpi)(\phidpim)$, $(\KSdK)(\phidpim)$,
 $(\phidpi)(\Kpipi)$, and $(\KSdK)(\Kpipi)$ invariant mass distributions.
By simultaneously fitting all four distributions, the normalization of the \Bzero
reflections in the $(\KSdK)(\phidpim)$ spectrum is constrained by the yields in the
high-statistics $(\phidpi)(\Kpipi)$ sample.
The components of the fit function for each invariant mass distribution are fully and partially 
reconstructed signals, reflections, and background.
The fully reconstructed \Bs and \Bzero signals are parametrized by the sum of two Gaussians
with relative normalizations and widths derived from simulation.
To account for discrepancies between data and simulation, a factor is introduced for the \Bs and \Bzero
signal shapes, respectively, that scales the widths of the Gaussians and that is allowed to float in the fit.
The shapes of partially reconstructed signal events and of reflections from
$\Bzero \rightarrow (\phidpi)(\Kpipi)$ mis-reconstructed as 
$\Bs \rightarrow (\phidpi)(\KSdKm)$ are determined from simulation using
empirical models.
Background from random combinations of tracks and other $B$ decays is described
by an exponential plus a constant function with all parameters floated in the fit.

The yield of fully reconstructed \Bzero mesons in the final state $i$, $(\phidpi)(\Kpipi)$
or $(\KSdK)(\Kpipi)$, is given by
\begin{eqnarray}
N_{\mathrm{rec},i}^{\Bzero} &=& N_\mathrm{tot}^{\Bzero} \Br{\BDDs} \Br{\DsKKpi} \nonumber \\
&& \cdot\, \Br{\DKpipi} \epsilon_i^{\Bzero},
\label{eqn:NBzero}
\end{eqnarray}
where $N_\mathrm{tot}^{\Bzero}$ is the total number of produced \Bzero mesons and is a free parameter in the fit,
the branching ratios are taken from Ref.~\cite{Nakamura:2010zzi}, and the efficiency $\epsilon_i^{\Bzero}$
is determined from simulation.
Equivalent expressions are used for the yields of partially reconstructed \Bzero decays
with an additional branching ratio factor for the \DSplus and \DsSplus decays.
The normalizations of reflections are calculated in the same way, but with the
efficiencies replaced by the mis-reconstruction fractions determined from simulation.
The number of fully reconstructed \Bs mesons in the final state $i$, $(\phidpi)(\phidpim)$ 
or $(\KSdK)(\phidpim)$, where the \Dsplus
decays in the same mode as the \Dsplus from the \Bzero decay is given by
\begin{equation}
N_{\mathrm{rec},i}^{\Bs} = N_{\mathrm{rec},i}^{\Bzero} \fDsDs \frac{\Br{\DsKKpi}}{\Br{\DKpipi}} 
\frac{\epsilon_i^{\Bs}}{\epsilon_i^{\Bzero}},
\end{equation}
with \fDsDs as a free parameter and $N_{\mathrm{rec},i}^{\Bzero}$ given by Eq. (\ref{eqn:NBzero}).
Equivalent equations hold for partially reconstructed \Bs decays.

\begin{figure}[htb]
\centering
\includegraphics[width=0.48\textwidth]{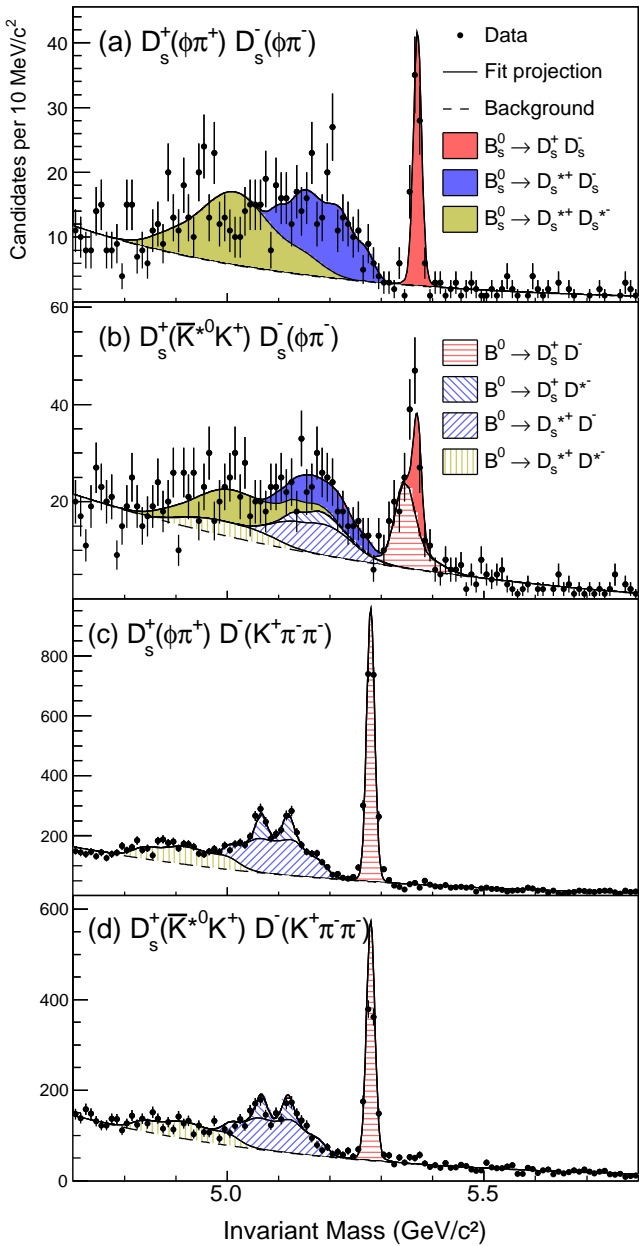}
\caption{Invariant mass distribution of
(a) $\Bs \rightarrow \Dsplus(\phidpi)\Dsminus(\phidpim)$,
(b) $\Bs \rightarrow \Dsplus(\KSdK)\Dsminus(\phidpim)$,
(c) $\Bzero \rightarrow \Dsplus(\phidpi)\Dminus(\Kpipi)$, and
(d) $\Bzero \rightarrow \Dsplus(\KSdK)\Dminus(\Kpipi)$ candidates
with the simultaneous fit projection overlaid.
The broader structures stem from decays where the photon or $\pi^0$ from the $D_{(s)}^{*+}$ decay is not reconstructed.
Misreconstructed signal events in (c) show up as reflections in (b).}
\label{fig:plots}
\end{figure}
Projections of the fit result are compared to the distribution of data events
in Fig.~\ref{fig:plots}.
The statistical significance of each signal exceeds 10$\,\sigma$ as estimated from a likelihood ratio of the fit 
with and without the signal component.

% systematics

% yields
% Signal parameterization
Systematic uncertainties on the fitted signal yields arise from the signal and background models.
Because the width scale factors of the fully reconstructed signal components are allowed to float in the fit,
the systematic uncertainties of these components are already included in the statistical errors.
To estimate the systematic effect due to the fixed shapes of the partially reconstructed signal components and reflections,
we repeat the fit multiple times with shape parameters
randomly varied according to the covariance matrix of the fits of the shapes to simulated data.
The mean deviations with respect to the central values are assigned as systematic uncertainties.
% Background parameterization
The systematic uncertainties due to the background mass model are estimated from
the changes in the results caused by using a second order polynomial instead of the sum of an exponential 
and a constant function.
% Optimization bias
By applying the selection optimization procedure on the normalization instead of the signal mode
we verified that a possible selection bias is negligible.

% efficiency
% Trigger
Systematic effects in the relative efficiency determination can be caused by a simulation that
does not describe the data accurately.
One source of systematic uncertainties is the trigger simulation, which can lead to a discrepancy 
in the $B$ meson momentum spectrum.
Although this effect cancels to first order in the ratio measurement,
it is accounted for by a reweighting of the simulated events.
The systematic uncertainties due to the detector simulation are estimated by the shift of the results
with respect to the case in which this reweighting is not applied.
% Lifetimes
The uncertainties on the world average \Bzero, \Dplus, and \Dsplus lifetimes are propagated 
by varying the lifetimes in the simulation.
For the \Bs lifetime, we consider two cases, the $1\sigma$ lower bound of the world average 
short-lived eigenstate lifetime and the $1\sigma$ upper bound of the mean \Bs lifetime.
% Dalitz model
The effects on the acceptance induced by variations of the \DsKKpi Dalitz structure are considered 
by generating different Dalitz model scenarios, with Dalitz model parameter values varied 
according to the systematic and correlated statistical uncertainties of the CLEO Dalitz fit.
The uncertainties of the \Dplus Dalitz model have a negligible effect on the result.
% P->VV
For \BsDsSDsS decays we investigate the effects of both a longitudinal polarization fraction \fL
deviating from our nominal assumption and a non-zero fraction of the \textit{CP}-odd component \fcpodd.
The fraction \fL is varied in the simulation according to the uncertainty of the \fL measurement
in $\Bzero \rightarrow \DSplus \DSminus$ decays~\cite{Aubert:2003jj}.
A variation of \fcpodd shows no effect on the \BsDsSDsS mass line shape, fit quality, 
or measured branching fraction ratios.
The effect of self cross-feed due to a wrong assignment of kaon and pion masses is negligible.

% External input: Intermediate and final state branching fractions
Further systematic uncertainties arise from external input quantities. 
The uncertainties of intermediate and final state branching fractions, \Br{\DsKKpi}, \Br{\DKpipi}, and \Br{\DSDsgammapi},
are propagated in the fit by adding Gaussian constraints to the corresponding fit parameters.
The resulting uncertainties of the measured branching fraction ratios are extracted by subtracting 
in quadrature the statistical uncertainties of the fits with branching fraction constrained and the
one where they are fixed to the central values.
When calculating the absolute branching fractions \Br{\BsDsDsInc} an additional relative uncertainty
of 16\,\% is introduced by the measurement uncertainties of $f_s/f_d$ and the branching fraction 
of the normalization channel \BDDs.
The systematic uncertainties are summarized in Table~\ref{tbl:overviewsys}.

\begin{table}
	\begin{tabular}{ccccc}
	\hline
	\hline
	Source&$\fDsDs$&$\fDsSDs$&$\fDsSDsS$&$\fDsDsInc$\\
	\hline
	Signal model		&$0.003$&$0.007$&$0.009$&$0.019$\\
	Background model	&$0.001$&$0.004$&$0.030$&$0.033$\\
	Detector simulation	&$0.001$&$0.003$&$0.010$&$0.005$\\
	$B$, $D$ lifetimes 	&${}^{+0.001}_{-0.002}$&${}^{+0.002}_{-0.004}$&${}^{+0.003}_{-0.006}$&${}^{+0.006}_{-0.012}$\\
	Dalitz model 		&$0.011$&$0.024$&$0.038$&$0.073$\\
	Helicity model 		&$0.001$&$0.005$&$0.012$&$0.008$\\
	Branching fractions	&$0.013$&$0.024$&$0.039$&$0.074$\\
	\hline
	Total                   &$0.017$&$0.035$&$0.065$&$0.112$ \\
	\hline
	\hline
	\end{tabular}
\caption[Overview of Systematic Uncertainties]
	{Overview of systematic uncertainties on the measured ratios of branching fractions.}
\label{tbl:overviewsys}
\end{table}

% result
%
As a result we obtain 
$\fDsDs = 0.183 \pm 0.021 \pm 0.017$,
$\fDsSDs = 0.424 \pm 0.046 \pm 0.035$,
$\fDsSDsS = 0.654 \pm 0.072 \pm 0.065$, and
$\fDsDsInc = 1.261 \pm 0.095 \pm 0.112$, 
where the first uncertainties are statistical and the second systematic.
Taking $\Br{\BDDs}= (7.2 \pm 0.8) \times 10^{-3}$ from Ref.~\cite{Nakamura:2010zzi} and $f_s/f_d = 0.269 \pm 0.033$ from 
Ref.~\cite{arXiv:0801.4375,Nakamura:2010zzi} an absolute inclusive
branching ratio of $\Br{\BsDsDsInc} = (3.38 \pm 0.25 \pm 0.30 \pm 0.56)$\,\% is
calculated where the third uncertainty comes from the normalization.
Assuming Eq.~(\ref{eqn:DGs}) to hold this would translate into a decay width difference
contribution of the \BsDsDsInc modes of 
$\DGs/\Gamma_s = (6.99 \pm 0.54 \pm 0.64 \pm 1.20)$\,\%, which is
consistent with the standard model expectation~\cite{Lenz:2006hd}.

% conclusions
In summary, we have measured the branching ratios of \BsDsDs, \BsDsSDs, \BsDsSDsS, and \BsDsDsInc decays 
relative to the normalization mode \BDDs.
Compared to previous analyses, we have reduced the systematic uncertainties by taking into account
the full \DsKKpi Dalitz structure, as opposed to using a simple two-body \Dsplus decay model.
The derived absolute branching ratios of
$\Br{\BsDsDs} = (0.49 \pm 0.06 \pm 0.05 \pm 0.08)$\,\%, $\Br{\BsDsSDs} = (1.13 \pm 0.12 \pm 0.09 \pm 0.19)$\,\%,
$\Br{\BsDsSDsS} = (1.75 \pm 0.19 \pm 0.17 \pm 0.29)$\,\%, and $\Br{\BsDsDsInc} = (3.38 \pm 0.25 \pm 0.30 \pm 0.56)$\,\%,
where the uncertainties are statistical, systematic, and due to the normalization,
are the most precise measurements to date.
The central values are lower than but consistent with the Belle result~\cite{Esen:2010jq} and
the previous CDF result, which is superseded by this measurement.

%===============================================================================
% ACKNOWLEDGEMENTS
%===============================================================================
We thank Mikhail S.~Dubrovin and David Cinabro for their help in implementing the CLEO Dalitz model.
We thank the Fermilab staff and the technical staffs of the participating institutions for their vital contributions. This work was supported by the U.S. Department of Energy and National Science Foundation; the Italian Istituto Nazionale di Fisica Nucleare; the Ministry of Education, Culture, Sports, Science and Technology of Japan; the Natural Sciences and Engineering Research Council of Canada; the National Science Council of the Republic of China; the Swiss National Science Foundation; the A.P. Sloan Foundation; the Bundesministerium f\"ur Bildung und Forschung, Germany; the Korean World Class University Program, the National Research Foundation of Korea; the Science and Technology Facilities Council and the Royal Society, UK; the Russian Foundation for Basic Research; the Ministerio de Ciencia e Innovaci\'{o}n, and Programa Consolider-Ingenio 2010, Spain; the Slovak R\&D Agency; the Academy of Finland; and the Australian Research Council (ARC). 
%\clearpage{}


\begin{thebibliography}{37}
\bibitem{Abazov:2011yk} 
  V.~M.~Abazov {\it et al.} (D0 Collaboration),
  Phys.\ Rev.\ D {\bf 84}, 052007 (2011).
\bibitem{ChargeConjugate}
  Charge-conjugate modes are implicitly included throughout this paper.
\bibitem{Aleksan:1993qp}
  R.~Aleksan {\it et al.},
  Phys.\ Lett.\ B {\bf 316}, 567 (1993).
\bibitem{Shifman:1987rj}
  M.~A.~Shifman and M.~B.~Voloshin,
  Yad.\ Fiz.\ {\bf 47}, 801 (1988) 
  [Sov.\ J.\ Nucl.\ Phys.\ {\bf 47}, 511 (1988)].
%\cite{Dunietz:2000cr}
\bibitem{Dunietz:2000cr} 
  I.~Dunietz, R.~Fleischer and U.~Nierste,
  Phys.\ Rev.\ D {\bf 63}, 114015 (2001).
\bibitem{Chua:2011er}
  C.~K.~Chua, W.~S.~Hou, and C.~H.~Shen,
  Phys.\ Rev.\  D {\bf 84}, 074037 (2011).
\bibitem{Aaltonen:2011nk}
  T.~Aaltonen {\it et al.} (CDF Collaboration),
  Phys.\ Rev.\ D {\bf 84}, 052012 (2011).
\bibitem{Barate:2000kd}
  R.~Barate {\it et al.} (ALEPH Collaboration),
  Phys.\ Lett.\ B {\bf 486}, 286 (2000).
\bibitem{Abulencia:2007zz}
  T.~Aaltonen {\it et al.} (CDF Collaboration),
  Phys.\ Rev.\ Lett.\  {\bf 100}, 021803 (2008).
\bibitem{Abazov:2008ig}
  V.~M.~Abazov {\it et al.} (D0 Collaboration),
  Phys.\ Rev.\ Lett.\  {\bf 102}, 091801 (2009).
\bibitem{Esen:2010jq}
  S.~Esen {\it et al.} (Belle Collaboration),
  Phys.\ Rev.\ Lett.\  {\bf 105}, 201802 (2010).
\bibitem{Nakamura:2010zzi}
  K.\ Nakamura \textit{et\ al.} (Particle Data Group), J.\ Phys. G \textbf {37}, 075021 (2010),
  and 2011 partial update.
\bibitem{arXiv:1110.2099}
  S.~Esen,
  arXiv:hep-ex/1110.2099.
\bibitem{Acosta:2004yw}
  D.~E.\ Acosta \textit {et~al.} (CDF Collaboration), Phys.\ Rev.\ D \textbf {71}, 032001 (2005).
\bibitem{Thomson:2002xp}%
  E.~J.\ Thomson \textit {et~al.}, IEEE Trans.\ Nucl.\ Sci., \textbf{49}, 1063 (2002);
%\bibitem{Ashmanskas:2003gf}
  B.~Ashmanskas {\it et al.},
  Nucl.\ Instrum.\ Methods A {\bf 518}, 532 (2004);
%
  L.~Ristori and G.~Punzi,
  Ann.\ Rev.\ Nucl.\ Part.\ Sci {\bf 60}, 595 (2010).
\bibitem{:2009tr}
  R.~E.~Mitchell {\it et al.} (CLEO Collaboration),
  Phys.\ Rev.\  D {\bf 79}, 072008 (2009).
\bibitem{Lange:2001uf}
  D.\ Lange, Nucl.\ Instrum.\ Methods A \textbf{462}, 152 (2001).
\bibitem{Aubert:2003jj}
  B.~Aubert {\it et al.} (BaBar Collaboration),
  Phys.\ Rev.\  D {\bf 67}, 092003 (2003).
\bibitem{Brun:1978fy}
  R.\ Brun, R.\ Hagelberg, M.\ Hansroul, and J.\ Lassalle, CERN-DD-78-2-REV, 1978 (unpublished). 
\bibitem{arXiv:0801.4375}
  T.~Aaltonen {\it et al.} (CDF Collaboration),
  Phys.\ Rev.\ D\ {\bf 77}, 072003 (2008).
\bibitem{Lenz:2006hd} 
  A.~Lenz and U.~Nierste,
  J.\ High Energy Phys.\ {\bf 0706}, 072 (2007); arXiv:1102.4274 [hep-ph].
\end{thebibliography}
\end{document}